\documentclass[aip,jap,graphicx,amsmath,amssymb,reprint,superscriptaddress,floatfix]{revtex4-1}
\usepackage{bm}
\usepackage{amsmath}
\usepackage{amssymb}
\usepackage{graphicx}
\usepackage{color}
\usepackage{soul,xcolor}
\setstcolor{red}
\usepackage{ulem}
\usepackage{hyperref}
\usepackage[utf8]{inputenc}
\usepackage[T1]{fontenc}
\usepackage{mathptmx}

\begin{document}

\title{Phenomenological analysis of transverse thermoelectric generation and cooling performance in magnetic/thermoelectric hybrid systems} 

\author{Kaoru Yamamoto}
\altaffiliation[Present Address: ]{NTT Secure Platform Laboratories, NTT Corporation, Musashino 180-8585, Japan}
\email{kaoru.yamamoto.uw@hco.ntt.co.jp}
\affiliation{Research Center for Magnetic and Spintronic Materials, National Institute for Materials Science (NIMS), Tsukuba 305-0047, Japan}

\author{Ryo Iguchi}
\affiliation{Research Center for Magnetic and Spintronic Materials, National Institute for Materials Science (NIMS), Tsukuba 305-0047, Japan}

\author{Asuka Miura}
\altaffiliation[Present Address: ]{Integrated Research for Energy and Environment Advanced Technology, Kyushu Institute of Technology, Kitakyushu, Fukuoka 804-8550, Japan}
\affiliation{Research Center for Magnetic and Spintronic Materials, National Institute for Materials Science (NIMS), Tsukuba 305-0047, Japan}

\author{Weinan Zhou}
\affiliation{Research Center for Magnetic and Spintronic Materials, National Institute for Materials Science (NIMS), Tsukuba 305-0047, Japan}

\author{Yuya Sakuraba}
\affiliation{Research Center for Magnetic and Spintronic Materials, National Institute for Materials Science (NIMS), Tsukuba 305-0047, Japan}
\affiliation{PRESTO, Japan Science and Technology Agency, Saitama 332-0012, Japan}

\author{Yoshio Miura}
\affiliation{Research Center for Magnetic and Spintronic Materials, National Institute for Materials Science (NIMS), Tsukuba 305-0047, Japan}
\affiliation{Center for Spintronics Research Network, Osaka University, Osaka 560-8531, Japan}

\author{Ken-ichi Uchida}
\email{UCHIDA.Kenichi@nims.go.jp}
\affiliation{Research Center for Magnetic and Spintronic Materials, National Institute for Materials Science (NIMS), Tsukuba 305-0047, Japan}
\affiliation{Institute for Materials Research, Tohoku University, Sendai 980-8577, Japan}
\affiliation{Center for Spintronics Research Network, Tohoku University, Sendai 980-8577, Japan}

\date{\today}

\begin{abstract}
We phenomenologically calculate the performance of the recently-observed Seebeck-driven transverse thermoelectric generation (STTG) for various systems in terms of the thermopower, power factor, and figure of merit to demonstrate the usefulness of STTG.
The STTG system consists of a closed circuit comprising thermoelectric and magnetic materials which exhibit the Seebeck and anomalous Hall effects, respectively. When a temperature gradient is applied to the hybrid system, the Seebeck effect in the thermoelectric material layer generates a longitudinal charge current in the closed circuit and the charge current subsequently drives the anomalous Hall effect in the magnetic material layer. 
The anomalous Hall voltage driven by the Seebeck effect has a similar symmetry to the transverse thermoelectric conversion based on the anomalous Nernst effect.
We find that the thermoelectric properties of STTG can be much better than those of the anomalous Nernst effect by increasing the Seebeck coefficient and anomalous Hall angle of the thermoelectric and magnetic materials, respectively, as well as by optimizing their dimensions.
We also formulate the electronic cooling performance in the STTG system, confirming the reciprocal relation for the hybrid transverse thermoelectric conversion. 
\end{abstract}


\maketitle 

\section{Introduction}
Thermoelectric power generation based on transverse thermoelectric phenomena has been actively investigated in spin caloritronics.\cite{Bauer2012NatMaterReview, Boona2014Ene&EngSciReview,Uchida2016ProcIEEESSEReview, Mizuguchi2019STAMreview,Uchida2021APL}
In particular, the anomalous Nernst effect (ANE), in which the thermoelectric voltage appears perpendicular to the direction of a temperature gradient and magnetization in magnetic materials, has gained much interest owing to its physical mechanism and thermoelectric functionalities.
ANE enables thermoelectric generation with a simple thermopile structure and a convenient scaling behavior.\cite{Sakuraba2013APExThermopile, Sakuraba2015ScrMaterANETEG, Yang2017AIPAdvCoiledWire}
By utilizing the features of ANE, a coil-shaped thermoelectric generator and a flexible heat flux sensor have recently been demonstrated.\cite{Yang2017AIPAdvCoiledWire,ZhouAPL2020heatflux}

For applications of the transverse thermoelectric generation, a large transverse thermopower of $>20 \ \mu\text{V }\text{K}^{-1}$ is at least necessary according to the estimation in Ref.~\onlinecite{Sakuraba2015ScrMaterANETEG}.
To realize such large transverse thermopower, ANE has been investigated in various materials including ferromagnetic alloys,\cite{Mizuguchi2012APEXFePtFilm,Hasegawa2015APLOrderedAlloyFilm, Isogami2017APExFe4Nfilm,Seki2018JphysDFePtfilm,Nakayama2019PRMateFeGa, Sakai2020NatureFe3Ga} Heusler compounds,\cite{Sakai2018NatPhysCo2MnGa, Reichlova2018APLCo2MnGafilm, Park2020PRBCMGthikcness,Sakuraba2020PRBCMAS,SakurabaCMG} and permanent magnets.\cite{Miura2019APLSmCo5,MiuraAPL2020Tempdep}
Multilayer films may produce large transverse thermoelectric voltage with increasing the number of layers.\cite{Uchida2015PRBMultiLayer,Fang2016PRBPtComultilayer,Ramos2019APLFe3O4/Pt,Seki2021PRB}
However, the obtained transverse thermopower is still much smaller than  $10 \ \mu\text{V }\text{K}^{-1}$; further materials exploration and device engineering are necessary to obtain larger transverse thermopower.

\begin{figure}
    \centering
    \includegraphics[width= 0.75\linewidth]{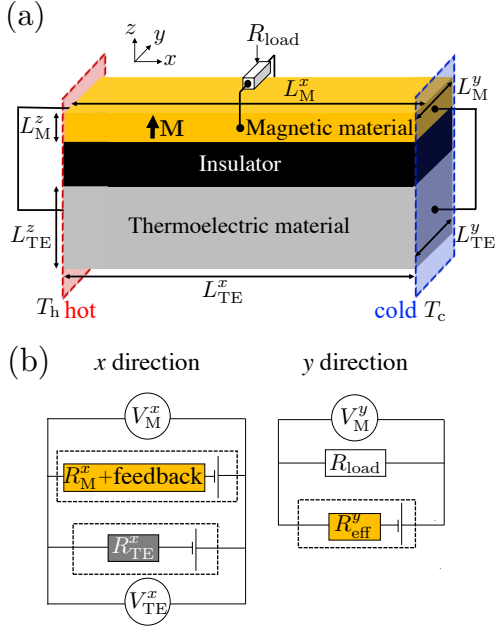}
    \caption{(a) A schematic illustration of the STTG system.
    (b) Equivalent circuits of the STTG system in the $x$ (left) and $y$ (right) directions.
    Battery symbols in the left circuit denote the electromotive force generated by the Seebeck effects in the thermoelectric and magnetic materials, while the symbol in the right circuit denotes the electromotive force generated by the ANE and Seebeck-driven AHE. 
    The internal resistance in the magnetic material in the $x$ direction includes the feedback effect due to AHE.}
   \label{fig:structure}
\end{figure}

Recently, Zhou \textit{et al}.~proposed and experimentally demonstrated a transverse thermoelectric generation with a similar symmetry to ANE but a different driving principle from ANE. \cite{Zou2020}
The device used in this experiment consists of a closed circuit comprising thermoelectric and magnetic materials which show the Seebeck effect and anomalous Hall effect (AHE), respectively.
When a temperature gradient is applied to the hybrid structure in the $x$ direction, the Seebeck effect in the thermoelectric material layer generates a charge current in the closed circuit and the charge current subsequently drives AHE in the magnetic material layer [see Fig.~\ref{fig:structure}(a)].
The anomalous Hall voltage driven by the Seebeck-effect-induced charge current has the same symmetry as the transverse thermoelectric conversion based on ANE when the magnetization $\mathbf{M}$ is along the $z$ direction, boosting the transverse thermopower in the magnetic layer.
Zhou \textit{et al}.~observed a giant transverse thermopower of 82.3 $\mu\text{V }\text{K}^{-1}$ in a $\text{Co}_2\text{MnGa}$/n-type Si hybrid structure, which is more than 10 times larger than the anomalous Nernst coefficient of a $\text{Co}_2\text{MnGa}$ monolayer.
The observed effect is referred to as the Seebeck-driven transverse thermoelectric generation (STTG).\cite{Zou2020}
They also demonstrated that the sign of the transverse thermopower induced by STTG can be changed by reversing the sign of the Seebeck coefficient of the thermoelectric material layer; a negative transverse thermopower of $-41.0$ $\mu\text{V }\text{K}^{-1}$ was observed in a $\text{Co}_2\text{MnGa}$/p-type Si structure.\cite{Zou2020}
The previous work, however, focused only on a few combinations of thermoelectric and magnetic materials.  

In this study, we phenomenologically calculate the thermopower, power factor, and figure of merit for STTG for various combinations of thermoelectric and magnetic materials to demonstrate the usefulness of STTG.
The STTG system can exhibit much better performance than existing ANE materials by optimizing transport properties and dimensions of the thermoelectric and magnetic materials.
Our results thus give strategies to realize efficient STTG-based energy harvesting and heat sensing devices. 
To confirm the reciprocal relation for the hybrid transverse thermoelectric conversion, we also formulate the electronic cooling performance in the STTG system. 

\begin{figure*}
    \centering
    \includegraphics[width= \linewidth]{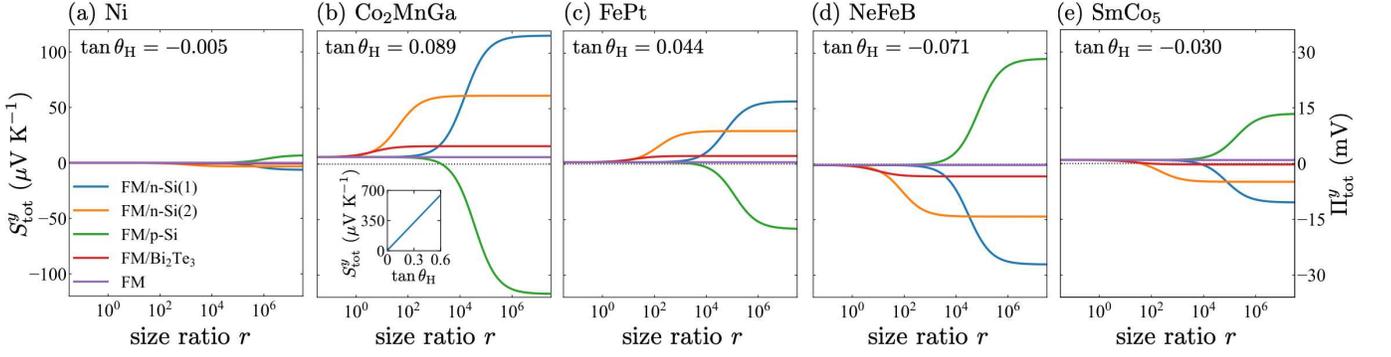}
    \caption{The size ratio $r$ dependence of the transverse thermopower $S_\text{tot}^y$ [Eq.~\eqref{eq:SANEeff}] for various ferromagnetic metals (FMs) [(a) Ni, (b) $\text{Co}_2\text{MnGa}$, (c) FePt, (d) NdFeB, and (e) $\text{SmCo}_5$] connected to typical thermoelectric materials (n-type Si(1), n-type Si(2), p-type Si, and $\text{Bi}_2\text{Te}_3$). 
    The inset to (b) shows the $\tan\theta_\text{H}$ dependence of $S_\text{tot}^y$ for the $\text{Co}_2\text{MnGa}$/n-Si(1) junction, where $\tan\theta_\text{H}$ is treated as a parameter with fixing $\rho_\text{M}$ at $r= 10^{5}$.}
   \label{fig:SANEeff}
\end{figure*}
\section{Model and setup}
The following phenomenological equations for the electric field $\mathbf{E}$ and heat current density $\mathbf{q}$ describes the transport properties in the STTG system:\cite{Harman1962JAPTheoryNernst,Landau,  Uchida2016ProcIEEESSEReview}
\begin{align}
\mathbf{E} &= \rho\mathbf{j}+S \nabla T + \rho_\text{AHE} (\mathbf{m}\times \mathbf{j})-S_\text{ANE} (\mathbf{m}\times \nabla T), \label{eq:E} \\
\mathbf{q} &= S T \mathbf{j}-\kappa\nabla T - S_\text{ANE}T (\mathbf{m}\times \mathbf{j}) + \kappa_\text{RL}(\mathbf{m}\times \nabla T), \label{eq:heat}
\end{align}
where $\mathbf{j}$, $\nabla T$, and $\mathbf{m}$ are the charge current density, the temperature gradient, and the unit vector along $\mathbf{M}$, respectively. 
$\rho = \rho_{xx}$ is the longitudinal resistivity, $S$ the Seebeck coefficient, $\kappa$ the thermal conductivity, $\rho_\text{AHE} = \rho_{yx}$ the anomalous Hall resistivity, $S_\text{ANE}$ the anomalous Nernst coefficient, and $\kappa_\text{RL}$ the Righi-Leduc coefficient.\cite{Zhang2000LeducRighi1, Li2017LeducRighi2}
Here, we neglect the thickness dependence of the transport coefficients and the magnetic field and magnetization dependences of $\rho$, $S$, and $\kappa$ for simplicity.
Note that this formalism can also be applied to the ordinary transverse transport phenomena by replacing $\mathbf{m}$ in Eqs.~\eqref{eq:E} and \eqref{eq:heat} with a magnetic field.
In our model calculation, we assume that the STTG system is in the isothermal condition in the $y$ direction, $\nabla_y T=0$,\cite{Harman1962JAPTheoryNernst,Uchida2016ProcIEEESSEReview}
and that charge and heat currents in the $z$ direction are perfectly blocked by an ideal insulator. 
Thus, the following calculation is independent of the dimensions of the insulator layer depicted in Fig.~\ref{fig:structure}(a). 
In practice, the insulator layer should be as thin as possible because a heat current in the layer does not contribute to the thermoelectric conversion.
In the following calculations, for simplicity, we neglect the interface effect at the junctions of each layer, although the interface effect may affect the transport properties when the thicknesses of the thermoelectric and magnetic layers are small.\cite{Uchida2015PRBMultiLayer,Fang2016PRBPtComultilayer,Ramos2019APLFe3O4/Pt,Seki2021PRB}

To determine the temperature gradients in the thermoelectric and magnetic materials, we solve $\nabla \cdot \mathbf{q} = \mathbf{E}\cdot \mathbf{j}$ with the following boundary conditions: $T_\text{TE(M)} = T_\text{h}$ at $x=0$ and $T_\text{TE(M)} = T_\text{c}$ at $x=L_\text{TE(M)}^x$. 
Here, $T_\text{TE(M)}$ is the temperature in the thermoelectric (magnetic) material, $T_\text{h(c)}$ is the temperature of the hot (cold) reservoir, and $L_\text{TE(M)}^x$ is the length of the thermoelectric (magnetic) material in the $x$ direction.
The notations for the dimensions along the $y$ and $z$ directions are defined in a similar manner.
We set $L_\text{TE}^x=L_\text{M}^x = L^x$ following the configuration depicted in Fig.~\ref{fig:structure}(a).
With the above conditions, we obtain
\begin{align}
\nabla_xT_\text{TE} &= -\frac{\Delta T}{L^x}+\frac{\rho_\text{TE}(j_\text{TE}^x)^2}{2\kappa_\text{TE}}(L^x-2x), \label{eq:nablaT_TE}  \\
\nabla_xT_\text{M} &=-\frac{\Delta T}{L^x}+\left\{\frac{\rho_\text{M}[(j_\text{M}^x)^2+(j_\text{M}^y)^2]}{2\kappa_\text{M}}\right. \notag\\
&\left.+\frac{\Delta T}{L^x}\frac{S_\text{ANE}j_\text{M}^y}{\kappa_\text{M}}\right\}(L^x-2x),  \label{eq:nablaT_M} 
\end{align}
where we assumed $\nabla_{x(y)}j_\text{M}^{x(y)} = 0$ and $S_\text{ANE}j_\text{M}^yL_\text{M}^x/\kappa_\text{M} \ll 1$.
Here, $\rho_\text{TE(M)}$, $\kappa_\text{TE(M)}$, and $j_\text{TE(M)}^x$ are the resistivity, the thermal conductivity, and the charge current density of the thermoelectric (magnetic) material, respectively, $j_\text{M}^y$ is the charge current density in the magnetic material in the $y$ direction, and $\Delta T = T_\text{h}-T_\text{c}$.

To calculate the thermoelectric performance in the STTG system, we need to solve Eq.~\eqref{eq:E} with the boundary conditions obtained from the equivalent circuits shown in Fig.~\ref{fig:structure}(b): $I_\text{TE}^x+I_\text{M}^x=0$, $V_\text{TE}^x = V_\text{M}^x$, and $V_\text{M}^y = R_\text{load}I_\text{M}^y$, where $I_{\text{TE}(\text{M})}^{x}$ and $I_{\text{TE}(\text{M})}^{y}$ are the charge currents in the thermoelectric (magnetic) material in the $x$ and $y$ directions, respectively,
$V_{\text{TE}(\text{M})}^{x} \equiv -\int_{0}^{L_{\text{TE}(\text{M})}^{x}}E_{\text{TE}(\text{M})}^{x}dx$ is the voltage in the thermoelectric (magnetic) material in the $x$ direction,  $V_{\text{M}}^{y} \equiv -\int_{0}^{L_\text{M}^{y}}E_{\text{M}}^{y}dy$ is the voltage in the magnetic material in the $y$ direction, $E_{\text{TE}(\text{M})}^{x}$ and $E_{\text{M}}^{y}$ are the electric fields, and $R_\text{load}$ is the load resistance.
In the following calculation, we use the value of $V_\text{M}^y$ at $x = L_\text{M}^x/2$.\cite{Harman1962JAPTheoryNernst}

\section{Results and discussions}
\subsection{Transverse thermopower}
By solving Eq.~\eqref{eq:E} with the aforementioned boundary conditions, we obtain the transverse thermopower for the STTG system as 
\begin{align}
    S_\text{tot}^y \equiv \left(\frac{E^y_\text{M}}{-\nabla_x T}\right)_{R_\text{load}\to\infty} = S_\text{ANE} + \frac{\rho_\text{AHE}}{\rho_\text{TE}/r+\rho_\text{M}}(S_\text{M}-S_\text{TE}), \label{eq:SANEeff}
\end{align}
where $S_\text{TE(M)}$ is the Seebeck coefficient of the thermoelectric (magnetic) material and $r \equiv L_\text{TE}^y L_\text{TE}^z/(L_\text{M}^yL_\text{M}^z)$ is the size ratio between the thermoelectric and magnetic materials; see Methods in Ref.~\onlinecite{Zou2020} for further details of the derivation.
Equation \eqref{eq:SANEeff} shows that the transverse thermopower for the STTG system can be enhanced owing to the superposition of the ANE contribution in the magnetic layer (first term) and the Seebeck-driven AHE: the STTG contribution (second term). 
Importantly, the second term can be designed by the combination of the thermoelectric and magnetic layers as well as their dimensions, i.e., $r$. 
With increasing $r$, the second term becomes effective and approaches $\tan\theta_\text{H}(S_\text{M}-S_\text{TE})$, where $\theta_\text{H} \equiv \tan^{-1}(\rho_\text{AHE}/\rho_\text{M})$ is the anomalous Hall angle of the magnetic layer.
Therefore, a large second term needs large values of $\theta_\text{H}$, $S_\text{TE}$, and $r$ because $S_\text{TE} \gg S_\text{M}$ for typical materials.

We demonstrate the behavior of $S_\text{tot}^y$ using the parameters of typical thermoelectric materials (n-type Si(1), n-type Si(2), p-type Si, and $\text{Bi}_2\text{Te}_3$) and magnetic materials (Ni, $\text{Co}_2\text{MnGa}$, $\text{L}1_0$-ordered FePt, NdFeB, and $\text{SmCo}_5$) shown in Table \ref{tab:Seebeck}, where n-type Si(1) and (2) have different transport properties and hereafter "type" is omitted.
$\text{Co}_2\text{MnGa}$ is a Heusler ferromagnet showing large AHE and ANE.\cite{Sakai2018NatPhysCo2MnGa, Reichlova2018APLCo2MnGafilm, SakurabaCMG},
$\text{L}1_0$-ordered FePt is often used in spintronics because it has a strong perpendicular magnetic anisotropy in a thin film form. \cite{Seki2008NmatFePt1, Seki2011APLFePt2}
NdFeB and $\text{SmCo}_5$ are rare-earth permanent magnets in practical use, which are known to exhibit substantially large AHE. \cite{Miura2019APLSmCo5}
In the previous work by Zhou \textit{et al.}, Ni, $\text{Co}_2\text{MnGa}$, and FePt were used for the experimental demonstration of STTG. \cite{Zou2020}
Figure \ref{fig:SANEeff}(a) [\ref{fig:SANEeff}(b)] shows $S_\text{tot}^y$ as a function of $r$ for the combination of the thermoelectric materials and Ni ($\text{Co}_2\text{MnGa}$).
In the case of Ni with negative $\tan\theta_\text{H}$, the second term of Eq.~\eqref{eq:SANEeff} is negative (positive) for n(p)-Si because of negative (positive) $S_\text{TE}$ and can be larger than $S_\text{ANE}$ owing to large $S_\text{TE}$ when $r$ is large.
We also find that the $r$ value at which the Seebeck-driven contribution is apparent with respect to $S_\text{ANE}$ for n-Si(1) and p-Si is larger than that for n-Si(2) and $\text{Bi}_2\text{Te}_3$ due to small $\rho_\text{TE}$ of n-Si(2) and $\text{Bi}_2\text{Te}_3$.
As shown in Fig.~\ref{fig:SANEeff}(b), the second term of Eq.~\eqref{eq:SANEeff} for $\text{Co}_2\text{MnGa}$ with positive $\tan\theta_\text{H}$ has the opposite sign to that for Ni for the same thermoelectric material. 
Importantly, when the magnetic material with large $\tan\theta_\text{H}$ is used, we can obtain the transverse thermopower of $>100\ \mu\text{V }\text{K}^{-1}$, which is more than an order of magnitude larger than $S_\text{ANE}$ of $\text{Co}_2\text{MnGa}$. 
The STTG contribution can further be increased in proportion to $\tan\theta_\text{H}$, as exemplified in the inset to Fig.~\ref{fig:SANEeff}(b).
In Figs.~\ref{fig:SANEeff}(c)-\ref{fig:SANEeff}(e), we show the behavior of $S_\text{tot}^y$ as a function of $r$ for $\text{L}1_0$-ordered FePt [Fig.~\ref{fig:SANEeff}(c)], NdFeB [Fig.~\ref{fig:SANEeff}(d)], and $\text{SmCo}_5$ [Fig.~\ref{fig:SANEeff}(e)], which enable STTG in the absence of external magnetic fields owing to their large coercive force and remanent magnetization.\cite{Miura2019APLSmCo5,Zou2020}
We find that NdFeB with relatively large $\tan\theta_\text{H}$ exhibits a larger STTG contribution than FePt and $\text{SmCo}_5$, although $|S_\text{ANE}|$ of NdFeB is smaller than that of FePt and $\text{SmCo}_5$.\cite{Miura2019APLSmCo5}
These demonstrations show that the transverse thermopower induced by STTG can be more than an order of magnitude larger than that induced by ANE by optimizing the combination of the thermoelectric and magnetic materials as well as their dimensions.

Here, we show that the insulator layer in Fig.~\ref{fig:structure} is important to obtain large STTG.
For the system in which thermoelectric and magnetic materials are directly connected in the $z$ direction, we can derive the transverse thermopower by solving Eq.~\eqref{eq:E} with the boundary condition based on the Maxwell’s equations: $E_\text{TE}^{x(y)} = E_\text{M}^{x(y)}$.
When the closed circuit is formed in the $x$ direction, the equivalent circuit in the $y$ direction gives $V_\text{M}^y = (I_\text{M}^y+I_\text{TE}^y)R_\text{load}$. 
The transverse thermopower in this case is calculated as 
\begin{equation}
S_\text{shunt}^y = \frac{1}{1+r\frac{\rho_\text{eff}^y}{\rho_\text{TE}}}S_\text{tot}^y,
\end{equation}
where we assume $L_\text{TE}^{x(y)} = L_\text{M}^{x(y)}$ and the linear temperature gradient along the $x$ direction for simplicity.
Here, $\rho_\text{eff}^y$ is the effective transverse resistivity of the STTG system \cite{Zou2020}:
\begin{equation}
    \rho_\text{eff}^y \equiv \rho_\text{M} + \frac{\rho_\text{AHE}^2}{\rho_\text{TE}/r+\rho_\text{M}} \label{eq:rhoeff}.
\end{equation}
Since the shunting factor $1/[1+r(\rho_\text{eff}^y/\rho_\text{TE})] \leq 1$, we obtain smaller transverse thermopower when the thermoelectric and magnetic materials are directly connected: $S_\text{shunt}^y \leq S_\text{tot}^y$.
In the following, therefore, we focus only on the case shown in Fig.~\ref{fig:structure}.

\subsection{Power factor}
We next discuss the power factor (PF) for the STTG system.
To derive PF,\cite{Zou2020} we first calculate the maximum output power generated by the transverse thermoelectric voltage, $P_\text{out}\equiv(V_\text{M}^y)^2/R_\text{load}$, with respect to the load resistance.
We calculate the maximum power $(P_\text{out})_\text{max}$ as follows:
\begin{equation}
(P_\text{out})_\text{max} =  \frac{L_\text{M}^yL_\text{M}^z}{L_\text{M}^x} \frac{({S_\text{tot}^y}\Delta T)^2}{4\rho_\text{eff}^y} \simeq L_\text{M}^xL_\text{M}^yL_\text{M}^z \frac{({S_\text{tot}^y}\nabla_x T)^2}{4\rho_\text{eff}^y}, \label{eq:power}
\end{equation}
where we used $\Delta T \simeq -L^x\nabla_x T$ with $\nabla_x T \simeq \nabla_x T_\text{TE} \simeq \nabla_x T_\text{M}$.
We then normalize $(P_\text{out})_\text{max}$ by the temperature gradient and the volume of the magnetic material and obtain PF for the STTG system as 
\begin{equation}
    \text{PF} \equiv \frac{{(S_\text{tot}^y)}^2}{\rho_\text{eff}^y}.\label{eq:PFeff}
\end{equation}
This expression becomes equivalent to the power factor for ANE, ${S_\text{ANE}}^2/\rho_\text{M}$, by taking the limit $r\to0$.
Since the second term of Eq.~\eqref{eq:rhoeff} is usually small, the parameter dependence of $\text{PF}$ is determined mainly by $S_\text{tot}^y$.
Although $\text{PF}$ in Eq.~\eqref{eq:PFeff} can be one or two orders of magnitude larger than that for ANE with appropriate choice and design of materials in a similar manner to $S_\text{tot}^y$, the difference between the volume of the magnetic material and the total volume of the STTG system should be taken into account to discuss the thermoelectric performance and efficiency of STTG.

\begin{table*}
\caption{\label{tab:Seebeck} Parameters of the thermoelectric and magnetic materials.}
\begin{ruledtabular}
\begin{tabular}{ccccccc}
Thermoelectric materials\cite{Qiao2019ACSBe2Te3, Zou2020} 
&  $S_\text{TE} \ (\mu\text{V }\text{K}^{-1})$ & $\kappa_\text{TE}\ (\text{W  }\text{m}^{-1}\ \text{K}^{-1})$  &$\rho_\text{TE} \ (\Omega \ \text{m})$   &  \\
\hline
n-Si(1) & $-1.3\times 10^{3}$ & $147$ & $4.1\times 10^{-2}$ \\
n-Si(2)& $-6.6\times 10^{2}$ & $121$ & $1.1\times 10^{-4}$  \\
p-Si & $1.3\times 10^{3}$ & $157$ & $9.9\times 10^{-2}$ \\
$\text{Bi}_2\text{Te}_3$  & $-1.5\times 10^{2}$ & $1$ & $1.3\times 10^{-5}$ \\
\hline
\hline
Magnetic materials\cite{Miura2019APLSmCo5,Miura2020PRM, Sakai2018NatPhysCo2MnGa, Zou2020}&  $S_\text{M} \ (\mu\text{V }\text{K}^{-1})$ & $\kappa_\text{M}\ (\text{W} \ \text{m}^{-1}\ \text{K}^{-1})$ &$\rho_\text{M} \ (\Omega \ \text{m})$ & $\rho_\text{AHE} \ (\Omega \ \text{m})$ & $\tan\theta_\text{H}$ & $S_\text{ANE} \ (\mu\text{V }\text{K}^{-1})$ \\
\hline 
Ni  & $-21.0$ & $76$ & $8.9 \times 10^{-8}$
& $-4.5\times 10^{-10}$ & $-0.005$ & $0.1$ \\
$\text{Co}_2\text{MnGa}$ & $-38.7$ & $22$ & $2.7\times 10^{-6}$  & $2.4\times 10^{-7}$ & $0.089$ & $6.3$ \\
FePt  & $-21.3$ & $30$ & $7.9\times 10^{-7}$
& $3.5\times 10^{-8}$ & $0.044$ & $1.7$ \\
NdFeB & $-5.7$ & $10$ & $1.4\times 10^{-6}$ & $-9.8\times10^{-8}$ & $-0.071$ & $-0.8$ \\
$\text{Sm}\text{Co}_{5}$  & $-18.9$ & $14$ & $5.5\times 10^{-7}$ & $-1.7\times 10^{-8}$ & $-0.030$ & $3.1$ \\
\end{tabular}
\end{ruledtabular}
\end{table*}

\subsection{Figure of merit}
Now, we are in a position to discuss the figure of merit for the transverse thermoelectric conversion in the STTG system. 
We maximize the efficiency $\eta \equiv P_\text{out}/(Q_\text{TE}^x|_{T=T_\text{h}}+Q_\text{M}^x|_{T=T_\text{h}})$ with respect to the load resistance $R_\text{load}$, where $Q_\text{TE(M)}^x|_{T=T_\text{h}}$ is the heat current from the hot reservoir in the thermoelectric (magnetic) material in the $x$ direction.
We then obtain the maximum efficiency as
\begin{equation}
    \eta_\text{max} = \eta_\text{c}\frac{1-\sqrt{1-Z_\text{tot}\overline{T}}}{1+\frac{T_\text{c}}{T_\text{h}}\sqrt{1-Z_\text{tot}\overline{T}}}, \label{eq:etamax}
\end{equation}
where $\eta_\text{c} = 1-T_\text{c}/T_\text{h}$ is the Carnot efficiency and $\overline{T} = (T_\text{h}+T_\text{c})/2$ is the average temperature of the hot and cold reservoirs.
Here
\begin{equation}
    Z_\text{tot}\overline{T} \equiv \frac{(S_\text{tot}^y)^2}{\rho_\text{eff}^y\kappa_\text{eff}^x}\overline{T} \label{eq:ZTeff}
\end{equation}
is the isothermal figure of merit for the STTG system, where
\begin{equation}
    \kappa_\text{eff}^x \equiv \kappa_\text{M} + r\kappa_\text{TE} + \frac{\overline{T}(S_\text{M}-S_\text{TE})^2}{\rho_\text{TE}/r + \rho_\text{M}} \label{eq:kappa}
\end{equation}
is the effective thermal conductivity of the STTG system with $\kappa_\text{TE(M)}$ being the thermal conductivity of the thermoelectric (magnetic) material.
Here, $Z_\text{tot}\overline{T} \leq 1$, which is a characteristic of the isothermal figure of merit for the transverse thermoelectric conversion.\cite{Harman1962JAPTheoryNernst}
In the same manner as $S_\text{tot}^y$ and $\text{PF}$, $Z_\text{tot}\overline{T}$ reduces to the isothermal figure of merit for ANE \cite{Harman1962JAPTheoryNernst} when $r\to0$.

Although $Z_\text{tot}\overline{T}$ is a complicated function of parameters of thermoelectric and magnetic materials, $Z_\text{tot}\overline{T}$ can be enhanced for a combination of a magnetic material with large $\tan\theta_\text{H}$ and a thermoelectric material with large figure of merit for the Seebeck effect, $Z_\text{TE}\overline{T} = {S_\text{TE}}^2\overline{T}/(\rho_\text{TE}\kappa_\text{TE})$, as discussed in Ref.~\onlinecite{Zou2020}.
This can be intuitively understood with naive approximations as follows.
The second term of $\rho_\text{eff}^y$ [Eq.~\eqref{eq:rhoeff}] and the third term of $\kappa_\text{eff}^x$ [Eq.~\eqref{eq:kappa}] are negligibly small for usual materials. 
In addition, we assume $\rho_\text{M}\kappa_\text{M} \ll \rho_\text{TE}\kappa_\text{TE}$ and consider only the STTG contribution in $S_\text{tot}^y$.
With these approximations, we find that the figure of merit for STTG takes its maximum of $Z_\text{tot}\overline{T} =Z_\text{TE}\overline{T}(\tan\theta_\text{H})^2/4$ at $r=\rho_\text{TE}/\rho_\text{M}$.
This approximation means that a maximum value of the figure of merit for STTG is mainly determined by $\tan\theta_\text{H}$ and $Z_\text{TE}\overline{T}$.

\begin{figure}
    \centering
    \includegraphics[width=0.85\linewidth]{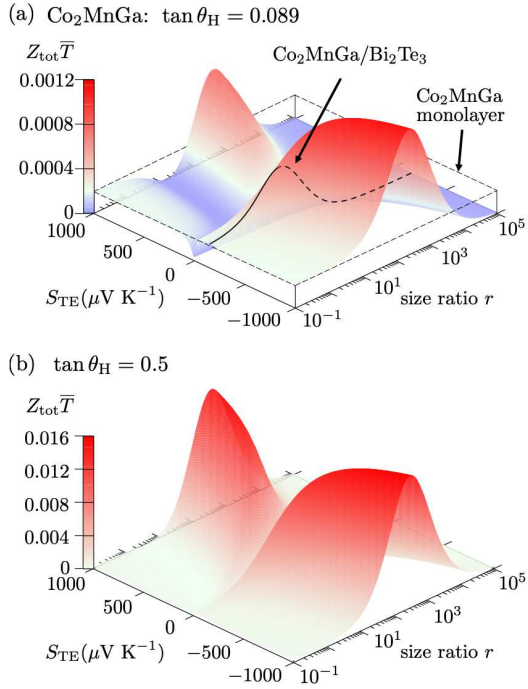}
    \caption{The figure of merit $Z_\text{tot}\overline{T}$ [Eq.~\eqref{eq:ZTeff}] for the STTG system at $\overline{T}=300\text{ K}$ as a function of $r$ and the Seebeck coefficient of the thermoelectric material $S_\text{TE}$. 
    In (a), the transport parameters of $\text{Co}_2\text{MnGa}$ are assumed. 
    In (b), $\tan\theta_\text{H}=0.5$ is assumed, where the transport parameters except for $\rho_\text{AHE}$ are fixed at the values for $\text{Co}_2\text{MnGa}$.  
    Although $S_\text{TE}$ is treated as a parameter, ${S_\text{TE}}^2/\rho_\text{TE}$ and $\kappa_\text{TE}$ are fixed at the values for $\text{Bi}_2\text{Te}_3$, i.e., $Z_\text{TE}\overline{T} = \text{const.}$, where the black curve in (a) shows the case for  $\text{Co}_2\text{MnGa}/\text{Bi}_2\text{Te}_3$ and the plane surrounded by dotted black lines shows the figure of merit for ANE in $\text{Co}_2\text{MnGa}$.
    The red (blue) color shows the larger (smaller) figure of merit than that for ANE in $\text{Co}_2\text{MnGa}$.}
   \label{fig:ZT}
\end{figure}

The behaviors of $Z_\text{tot}\overline{T}$ are exemplified in Fig.~\ref{fig:ZT} (note that the aforementioned approximations are not applied here).
Figure \ref{fig:ZT}(a) shows $Z_\text{tot}\overline{T}$ for $\text{Co}_2\text{MnGa}$ at $\overline{T}=300\text{ K}$ as a function of $r$ and $S_\text{TE}$.
Although $S_\text{TE}$ is treated as a parameter, ${S_\text{TE}}^2/\rho_\text{TE}$ and $\kappa_\text{TE}$ are fixed at the values for $\text{Bi}_2\text{Te}_3$, i.e., $Z_\text{TE}\overline{T} = \text{const.}$, where the black curve in Fig.~\ref{fig:ZT}(a) shows the case for the $\text{Co}_2\text{MnGa}$/$\text{Bi}_2\text{Te}_3$ junction with the parameters shown in Table \ref{tab:Seebeck}.
By choosing optimal $r$, $Z_\text{tot}\overline{T}$ can be larger than the figure of merit for ANE in $\text{Co}_2\text{MnGa}$ both for positive and negative $S_\text{TE}$ values.
The difference in the maximum $Z_\text{tot}\overline{T}$ between positive and negative $S_\text{TE}$ values is attributed to the $S_\text{ANE}$ offset in Eq.~\eqref{eq:SANEeff}; the situation depends on the sign and magnitude of $S_\text{ANE}$. 
As shown in  Fig.~\ref{fig:ZT}(b), when $\tan\theta_\text{H} = 0.5$ is assumed, the  maximum value of $Z_\text{tot}\overline{T}$ is dramatically improved in comparison with the figure of merit for ANE in $\text{Co}_2\text{MnGa}$.
Here, due to the dominant contribution of STTG, the significant improvement of $Z_\text{tot}\overline{T}$ appears both for positive and negative $S_\text{TE}$ values.
This demonstration shows that not only $S_\text{tot}^y$ and $\text{PF}$ but also $Z_\text{tot}\overline{T}$ can be enhanced by AHE driven by the Seebeck effect.

\subsection{Thermoelectric cooling}
Finally, we mention that the STTG system also works as a transverse thermoelectric temperature modulator by replacing the load resistance with an external battery in Fig.~\ref{fig:structure}.
The external battery induces a charge current in the $y$ direction, and the charge current is bent in the $x$ direction by AHE.
When the charge current in the $x$ direction flows in the closed circuit comprising the thermoelectric and magnetic materials, heat is generated or absorbed at the junctions by the Peltier effect.
This is the reciprocal process of STTG.
The resultant temperature gradient under the adiabatic condition in the $x$ direction is calculated as follows.
Since we use the same boundary conditions for $T_\text{TE(M)}$, the temperature gradients in the thermoelectric and magnetic materials are the same as Eqs.~\eqref{eq:nablaT_TE} and \eqref{eq:nablaT_M}, respectively.
By solving Eq.~\eqref{eq:E} with the boundary conditions $I_\text{TE}^x+I_\text{M}^x =0$, $V_\text{TE}^x = V_\text{M}^x$, we obtain $I_\text{M}^x$ and $I_\text{M}^y$ as follows:
\begin{align}
I^x_\text{M} &= \frac{S_\text{M}-S_\text{TE}}{R_\text{TE}^x+R_\text{M}^x}\Delta T +  \frac{L_\text{M}^x}{L_\text{M}^y}\frac{R_\text{AHE}}{R_\text{TE}^x+R_\text{M}^x} I_\text{M}^y, \label{eq:I_M^x} \\
I_\text{M}^y &= -\frac{V_\text{M}^y}{R_\text{eff}^y} -\frac{L_\text{M}^y}{L_\text{M}^x} \frac{S_\text{tot}^y}{R_\text{eff}^y}\Delta T, \label{eq:I_M^y} 
\end{align}
where $R_\text{TE(M)}^x = \rho_\text{TE(M)}L_\text{TE(M)}^x/(L_\text{TE(M)}^yL_\text{TE(M)}^z)$ is the resistance of the thermoelectric (magnetic) material in the $x$ direction, $R_\text{AHE} = \rho_\text{AHE}L_\text{M}^y/(L_\text{M}^zL_\text{M}^x)$ is the anomalous Hall resistance, and $R_\text{eff}^y = \rho_\text{eff}^yL_\text{M}^y/(L_\text{M}^zL_\text{M}^x)$ is the effective resistance in the magnetic material in the $y$ direction. 
Using Eq.~\eqref{eq:I_M^x} and $Q_\text{TE}^x + Q_\text{M}^x = 0$, we obtain the resultant temperature gradient under the adiabatic condition in the $x$ direction as follows:
\begin{equation}
    \nabla_x T = \frac{\Pi_\text{tot}^yE_\text{M}^y}{\kappa_\text{eff}^x\rho_\text{eff}^y(1-Z_\text{tot}T)} \simeq \frac{\Pi_\text{tot}^yE_\text{M}^y}{\kappa_\text{eff}^x\rho_\text{eff}^y}, \label{eq:tempgrad}
\end{equation}  where we assumed $Z_\text{tot}T \ll 1$ and the linear temperature gradient in the $x$ direction for simplicity.
Here, $\Pi_\text{tot}^y$ is the transverse charge-to-heat conversion coefficient for the STTG system defined as
\begin{equation}
\Pi_\text{tot}^y =  \left(\frac{q_\text{M}^x+rq_\text{TE}^x}{j_\text{M}^y}\right)_{\nabla_x T=0}, \label{eq:Pitot}
\end{equation}
where $q_\text{TE(M)}^x$ is the heat current density in the thermoelectric (magnetic) material in the $x$ direction.
We note that, in the definition of $\Pi_\text{tot}^y$, the aforementioned adiabatic condition in the $x$ direction is not assumed.
The coefficient satisfies the reciprocal relation $\Pi_\text{tot}^y = TS_\text{tot}^y$ (see the right longitudinal axis in Fig.~\ref{fig:SANEeff}).
By taking the limit $r\to 0$, $\Pi_\text{tot}^y$ reduces to the anomalous Ettingshausen coefficient, $\Pi_\text{AEE} = TS_\text{ANE}$, and
the temperature gradient in Eq.~\eqref{eq:tempgrad} reduces to that for the anomalous Ettingshausen effect.\cite{Seki2018JphysDFePtfilm,Seki2018APL}

\begin{figure}
    \centering
    \includegraphics[width=\linewidth]{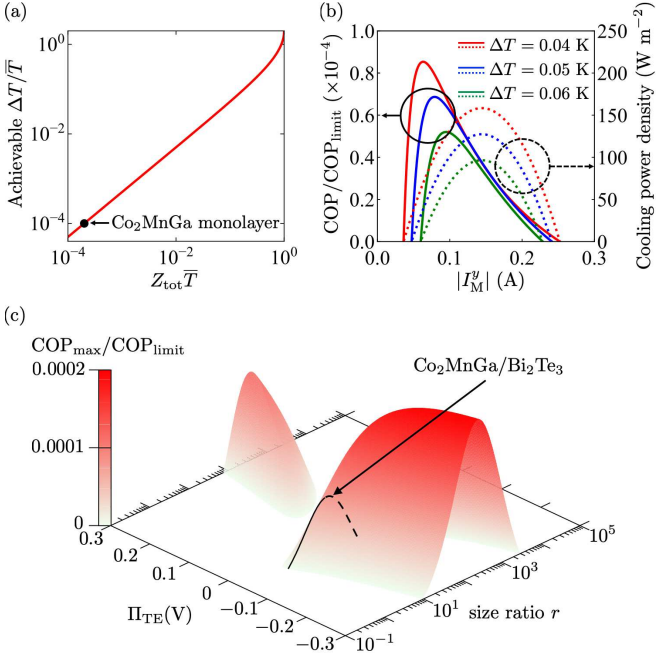}
    \caption{(a) The achievable $\Delta T/\overline{T}$ [Eq.~\eqref{eq:maxT}] as a function of $Z_\text{tot}\overline{T}$. The black circle shows the achievable $\Delta T/\overline{T}$ for the figure of merit for ANE in $\text{Co}_2\text{MnGa}$. 
    (b) $\text{COP}$ [Eq.~\eqref{eq:COP}] normalized by $\text{COP}_\text{limit}$ and the cooling power density for $\text{Co}_2\text{MnGa}/\text{Bi}_2\text{Te}_3$ at $\overline{T}=300\text{ K}$ as a function of $|I_\text{M}^y|$, where we set $L_\text{TE(M)}^x = L_\text{TE(M)}^y = 10\text{ mm}$, $L_\text{M}^z = 0.1 \text{ mm}$, and $L_\text{TE}^z = 1\text{ mm}$.
    (c) The maximum COP [Eq.~\eqref{eq:COPmax}] normalized by $\text{COP}_\text{limit}$ at $\overline{T}=300\text{ K}$ and $\Delta T = 0.05 \text{ K}$ as a function of $r$ and the Peltier coefficient of the thermoelectric material $\Pi_\text{TE}$, where the parameters are set in the same manner as Fig.~\ref{fig:ZT}(a).  
    The black curve in (c) shows the case for  $\text{Co}_2\text{MnGa}/\text{Bi}_2\text{Te}_3$.
    We only plot the region where $\Delta T = 0.05 \text{ K}$ is achievable.
    }
   \label{fig:COP}
\end{figure}

We define the coefficient of performance (COP) for the STTG system as
\begin{align}
\text{COP} &= -\frac{Q_\text{TE}^x|_{T=T_\text{c}} + Q_\text{M}^x|_{T=T_\text{c}}}{I_\text{M}^y V_\text{ext}} \notag \\
&=-\frac{\frac{R_\text{eff}^y}{2}(I_\text{M}^y)^2+\frac{L_\text{M}^y}{L_\text{M}^x}T_\text{h}S_\text{tot}^yI_\text{M}^y+K_\text{eff}^x\Delta T}{R_\text{eff}^y(I_\text{M}^y)^2 +\frac{L_\text{M}^y}{L_\text{M}^x}S_\text{tot}^y \Delta T I_\text{M}^y}, \label{eq:COP} 
\end{align}
where $V_\text{ext} = -V_\text{M}^y$ is the voltage of the external battery and $K_\text{eff}^x = \kappa_\text{eff}^x L_\text{M}^yL_\text{M}^z/L_\text{M}^x$ is the effective thermal conductance of the STTG device.
Here, we used Eqs.~\eqref{eq:heat}-\eqref{eq:nablaT_M} and \eqref{eq:I_M^x} for the numerator, and Eq.~\eqref{eq:I_M^y} for the denominator.
To cool the STTG system, the heat current should be absorbed from the cold reservoir, $Q_\text{TE}^x|_{T=T_\text{c}} + Q_\text{M}^x|_{T=T_\text{c}} \leq 0$, and this gives the following condition for the temperature difference:
\begin{equation}
\frac{\Delta T}{\overline{T}} \leq \frac{2}{Z_\text{tot}\overline{T}} \left(2-Z_\text{tot}\overline{T}-2\sqrt{1-Z_\text{tot}\overline{T}}\right). \label{eq:maxT}
\end{equation}
We show the upper bound of $\Delta T/\overline{T}$ in Fig.~\ref{fig:COP}(a); this is the achievable $\Delta T/\overline{T}$ for a given $Z_\text{tot}\overline{T}$.
Figure \ref{fig:COP}(b) shows the $|I_\text{M}^y|$ dependence of COP and the cooling power density defined as $-(Q_\text{TE}^x|_{T=T_\text{c}} + Q_\text{M}^x|_{T=T_\text{c}})/L_\text{M}^yL_\text{M}^z$.
The behaviors are similar to those for the conventional Peltier and Ettingshausen effects.\cite{Mobarak2021PRAppl}
The maximum COP with respect to $I_\text{M}^y$ is calculated to be
\begin{equation}
\text{COP}_\text{max} =\text{COP}_\text{limit}\frac{1-\frac{T_\text{h}}{T_\text{c}}\sqrt{1-Z_\text{tot}\overline{T}}}{1+\sqrt{1-Z_\text{tot}\overline{T}}}, \label{eq:COPmax}
\end{equation}
where $\text{COP}_\text{limit} = T_\text{c}/\Delta T$ is the achievable limit of COP.
Equation \eqref{eq:COPmax} reduces to the maximum COP for the anomalous Ettingshausen effect\cite{Harman1962JAPCOP} when $r\to0$. 
$\text{COP}_\text{max}$ for the STTG system can be larger than that for the anomalous Ettingshausen effect alone; its behavior is similar to that of $Z_\text{tot}\overline{T}$ [compare Fig.~\ref{fig:COP}(c) with \ref{fig:ZT}(a)]. 

\section{Conclusion}
We have phenomenologically calculated the thermoelectric properties of STTG for various combinations of thermoelectric and magnetic materials and demonstrated its usefulness.
We have shown that, by combining a thermoelectric material with a large Seebeck coefficient and a magnetic material with a large anomalous Hall angle, not only the transverse thermopower but also the figure of merit for STTG can be much larger than those for conventional ANE.
We have also discussed the reciprocal process of STTG; the combination of AHE in a magnetic material and the Peltier effect in a thermoelectric material enables transverse charge-to-heat conversion. 
It is worth mentioning that STTG works even if the thermoelectric material is also ferromagnetic, although the formulation becomes complicated. Thus, the experimental demonstration of STTG in all-ferromagnetic systems is desired.
The phenomenological model for STTG will invigorate materials science research and device engineering in spin caloritronics, paving the way for versatile energy harvesting, heat sensing, and thermal management applications. 

\begin{acknowledgments}
This work was supported by CREST “Creation of Innovative Core Technologies for Nano-enabled Thermal Management” (Grant No.~JPMJCR17I1) and PRESTO “Scientific Innovation for Energy Harvesting Technology” (Grant No.~JPMJPR17R5) from JST, Japan, and Mitou challenge 2050 (Grant No.~P14004) from NEDO, Japan. 
A.~M.~was supported by JSPS through Research Fellowship for Young Scientists (Grant No.~JP18J02115). 
\end{acknowledgments}

\section*{Data Availability}
The data that support the findings of this study are available from the corresponding author upon reasonable request.

%

\end{document}